\documentclass[aps,pra,twocolumn,grouped address,notitlepage]{revtex4-2}

\usepackage{amsmath,amssymb,amsfonts}
\usepackage{graphicx}
\usepackage{natbib}
\usepackage[normalem]{ulem}
\usepackage{xcolor}
\usepackage[caption=false,labelformat=simple]{subfig} 

\usepackage{url}
\usepackage[breaklinks]{hyperref}
\hypersetup{
    unicode=true,
    colorlinks=true,
    linkcolor=blue,
    citecolor=blue,
    urlcolor=blue
  }
	\usepackage{breakurl}

\begin{document}

\title{Effective anisotropic interaction potentials \\
for pairs of ultracold molecules shielded by a static electric field}
\author{Bijit Mukherjee}
\email{bijit.9791@gmail.com}
\affiliation{Faculty of Physics, University of Warsaw, Pasteura 5, 02-093 Warsaw, Poland}
\author{Luis Santos}
\affiliation{Institut f\"ur Theoretische Physik, Leibniz Universit\"at Hannover, Appelstrasse 2, D-30167 Hannover, Germany}
\author{Jeremy M. Hutson}
\email{j.m.hutson@durham.ac.uk} \affiliation{Joint Quantum Centre (JQC)
Durham-Newcastle, Department of Chemistry, Durham University, South Road,
Durham, DH1 3LE, United Kingdom.}

\begin{abstract}
Quantum gases of ultracold polar molecules have novel properties because of the strong dipolar forces between molecules. Current experiments shield the molecules from destructive collisions by engineering long-range repulsive interactions using microwave or static electric fields. These shielding methods produce interaction potentials with large repulsive cores that are not well described with contact potentials. In this paper we explore the anisotropic interaction potentials that arise for pairs of polar molecules shielded with static electric fields. We derive computationally inexpensive approximations for the potentials that are suitable for use in calculations of many-body properties. The interaction potentials for molecules shielded with static fields are substantially different from those that arise from microwave shielding and will produce quite different many-body physics.
\end{abstract}

\date{\today}

\maketitle

\section{Introduction}

Recent advances in experiment and theory for ultracold molecules have led to the production of Fermi-degenerate gases \cite{DeMarco:KRb-degen:2019, Valtolina:KRb2D:2020, Schindewolf:NaK-degen:2022, Duda:NaK-Feshback-degen:2023} and Bose-Einstein condensates \cite{Bigagli:BEC:NaCs:2024} of polar molecules. These advances open new frontiers in the physics of dipolar gases. Molecules with permanent electric dipole moments can produce much stronger dipole-dipole interactions than magnetic atoms \cite{Chomaz:2022}. Molecular dipoles are generally weaker than those of Rydberg atoms \cite{Browaeys:2020}, but ultracold polar molecules can have long lifetimes and coherence times \cite{Ruttley:magic:2025} that are hard to achieve for Rydbergs.

Left to themselves, ultracold polar molecules do not stay trapped for long. They undergo collisions with one another that result in trap loss. The loss is not fully understood \cite{Bause:2023}, but is believed to occur via formation of 2-molecule collision complexes \cite{Mayle:2013} followed by excitation due to the trapping laser \cite{Christianen:laser:2019}. In any case, the loss can be effectively prevented by \emph{shielding} methods, which engineer long-range potential barriers to prevent colliding pairs reaching short range, where the losses occur.

There are two types of shielding in current use. Static-field shielding \cite{Avdeenkov:2006, Wang:dipolar:2015, Gonzalez-Martinez:adim:2017, Li:KRb-shield-3D:2021, Mukherjee:CaF:2023, Mukherjee:alkali:2024} engineers the repulsive barrier by using a strong electric field to tune close to a 2-molecule degeneracy, and the dipole-dipole interaction then produces a repulsive barrier in the effective potential for the shielded state. Microwave shielding \cite{Karman:shielding:2018, Lassabliere:2018, Karman:shielding-imp:2019, Karman:ellip:2020, Anderegg:2021, Karman:res:2022, Deng:microwave:2023} engineers an analogous repulsive barrier by using microwave radiation, blue-detuned from a molecular rotational transition, to create a similar near-degeneracy.

For either type of shielding, the interaction between molecules is characterized by an effective interaction potential that is very different from that for magnetic atoms, and also different from that assumed in most theories of dipolar gases. The major difference is that, for magnetic atoms, repulsion occurs at very short range (around 10 bohr). Such repulsive interactions are fairly well represented with zero-range contact potentials modeled with Dirac delta-functions. This is the formulation used, for example, in the widely used theoretical \emph{ansatz} of Yi and You \cite{Yi:2000, Yi:2001}. However, the interactions between shielded molecules are characterized by a very large repulsive core. The excluded volume corresponds to a range of order 1000 bohr \cite{Wang:dipolar:2015, Karman:shielding:2018, Mukherjee:alkali:2024}. For the large repulsive cores encountered for shielded molecules, a contact potential is a very poor approximation. This is likely to produce very different many-body physics \cite{Langen:dipolar-droplets:2025, Jin:Bose:2025, Ciardi:self:2025}.

To simulate many-body properties of systems made up of shielded ultracold molecules, we need simple representations of the effective potentials involved. For molecules shielded by a circularly polarised microwave field \cite{Karman:shielding:2018}, suitable potentials have been provided by Deng \emph{et al.}\ \cite{Deng:microwave:2023} and used for many-body simulations \cite{Langen:dipolar-droplets:2025, Jin:Bose:2025}. In this case the potentials are antidipolar at long range, meaning that they are repulsive for end-to-end approach of two molecules and attractive for side-to-side approach. At shorter range (inside a few hundred bohr), there is a repulsive potential that is itself strongly anisotropic.

For molecules shielded by static fields, the long-range potentials are qualitatively different from those for {circularly polarised} microwaves. They are similar to those for dipoles fixed in space, and are attractive for end-to-end approach and repulsive for side-to-side approach. Nevertheless, the effective potentials still have large repulsive cores \cite{Mukherjee:CaF:2023, Mukherjee:alkali:2024} and these are highly anisotropic. Similar topologies can be achieved for microwave shielding by supplementing the circularly polarised field with a linearly polarised one \cite{Karman:double:2025, Deng:double-microwave:2025}. Such topologies are likely to lead to very different behaviour for many-body systems. In this paper we investigate the anisotropic interaction potentials for molecules shielded by static fields and derive computationally inexpensive approximate forms for them that are suitable for quantum many-body calculations.

\section{Theory}

\subsection{Hamiltonian and its matrix elements} \label{ssec:ham}

We follow a theoretical approach similar to that in ref.\ \cite{Mukherjee:CaF:2023}. The Hamiltonian for a single polar molecule in a $^1\Sigma$ state, in the presence of an electric field $\boldsymbol{F}$, is
\begin{equation}
\hat{h} = b\hat{\boldsymbol{n}}^2 - \boldsymbol{\mu} \cdot \boldsymbol{F},
\label{eq:ham-Stark}
\end{equation}
where $\hat{\boldsymbol{n}}$ is the operator for molecular rotation, $b$ is the rotational constant, and $\boldsymbol{\mu}$ is the dipole moment along the molecular axis.

The Hamiltonian for a colliding pair of molecules is
\begin{equation}
\hat{H} = -\frac{\hbar^2}{2\mu_\textrm{red}} \hat{\boldsymbol{\nabla}}^2 + \hat{h}_1 + \hat{h}_2  + \hat{H}_\textrm{dd},
\label{eq:ham-pair}
\end{equation}
where $\mu_\textrm{red}$ is the reduced mass, $\hat{\boldsymbol{\nabla}}^2$ is the Laplacian, $\hat{h}_k$ is the Hamiltonian of molecule $k$ from Eq.\ \ref{eq:ham-Stark}, and $\hat{H}_\textrm{dd}$ represents the dipole-dipole interaction between the pair of molecules.

The dipole-dipole operator is
\begin{equation}
\hat{H}_\textrm{dd} = \frac{\boldsymbol{\mu}_1 \cdot \boldsymbol{\mu}_2 - 3(\boldsymbol{\mu}_1 \cdot \hat{\boldsymbol{R}}) (\boldsymbol{\mu}_2 \cdot \hat{\boldsymbol{R}})}{4\pi\epsilon_0 R^3},
\label{eq:V_dd}
\end{equation}
where $\boldsymbol{\mu}_k$ is the dipole moment of molecule $k$ and $\boldsymbol{R}$ is the intermolecular vector, with corresponding unit vector $\hat{\boldsymbol{R}}$ and length $R$. In a spherical tensor representation,
\begin{equation}
\hat{H}_\textrm{dd}  = -\sqrt{6} \frac{\mu_1\mu_2} {4\pi\epsilon_0 R^3} \sum_p (-1)^p \left[ \hat{\boldsymbol{r}}_1 \otimes \hat{\boldsymbol{r}}_2 \right]_{2,p} C_{2,-p} (\theta, \phi),
\end{equation}
where $\hat{\boldsymbol{r}}_k$ is a unit vector along the axis of molecule $k$.
Here, the functions $C_{k,p}(\theta,\phi)$ are the Racah-normalised spherical harmonics, and $\theta$ and $\phi$ indicate the orientation of $\boldsymbol{\hat{R}}$ with respect to the laboratory-frame axis $Z$. The tensor component $\left[ \hat{\boldsymbol{r}}_1 \otimes \hat{\boldsymbol{r}}_2 \right]_{2,p}$ is
\begin{align}
    \left[ \hat{\boldsymbol{r}}_1 \otimes \hat{\boldsymbol{r}}_2 \right]_{2,p} &= \sqrt{5} \sum_{p_1,p_2} (-1)^p
    \begin{pmatrix}
    2 & 1 & 1 \\
    p & -p_1 & -p_2
    \end{pmatrix} \nonumber \\
    & \times
    C_{1,p_1} (\beta_1,\alpha_1)
    C_{1,p_2} (\beta_2,\alpha_2).
\end{align}
The angles $(\beta_k,\alpha_k)$ indicate the orientation of $\hat{\boldsymbol{r}}_k$ in the laboratory frame.

In a basis set of spherical harmonics for two freely rotating molecules, $|n_1, m_1; n_2,m_2\rangle$, where $m$ is the projection of $n$ onto the space-fixed axis $Z$, the matrix elements of $\hat{H}_\textrm{dd}$ are
\begin{align}
& \langle n_1, m_1; n_2, m_2 | \hat{H}_\textrm{dd} | n^{\prime}_1, m^{\prime}_1;
n^{\prime}_2, m^{\prime}_2 \rangle \nonumber \\
&= -\sqrt{30} \frac{\mu_1 \mu_2}{4\pi\epsilon_0 R^3}
    \begin{pmatrix}
        2 & 1 & 1 \\
        p & -p_1 & -p_2
    \end{pmatrix}
    \langle n_1, m_1 | C_{1,p_1} | n^{\prime}_1, m^{\prime}_1 \rangle \nonumber \\
    &\times
    \langle n_2, m_2 | C_{1,p_2} | n^{\prime}_2, m^{\prime}_2 \rangle  C_{2,-p}(\theta, \phi),
\label{eq:v_dd_basis1}
\end{align}
where $p_1=m_1-m^{\prime}_1$, $p_2=m_2-m^{\prime}_2$ and $p=p_1+p_2$.

We expand Eq.\ (\ref{eq:v_dd_basis1}) to obtain
\begin{align}
& \langle n_1, m_1; n_2, m_2 | \hat{H}_\textrm{dd} | n^{\prime}_1, m^{\prime}_1;
n^{\prime}_2, m^{\prime}_2 \rangle \nonumber \\
&= (-1)^{1+m_1+m_2}\sqrt{30} \frac{\mu_1 \mu_2}{4\pi\epsilon_0 R^3}
    \begin{pmatrix}
        2 & 1 & 1 \\
        p & -p_1 & -p_2
    \end{pmatrix}C_{2,-p}(\theta, \phi) \nonumber \\
& \times \sqrt{(2n_1+1)(2n^{\prime}_1+1)}    \begin{pmatrix}
       n_1 & 1 & n^{\prime}_1 \\
       -m_1 & p_1 & m^{\prime}_1
   \end{pmatrix}
     \begin{pmatrix}
       n_1 & 1 & n^{\prime}_1 \\
      0 & 0 & 0
   \end{pmatrix} \nonumber \\
   & \times \sqrt{(2n_2+1)(2n^{\prime}_2+1)} \begin{pmatrix}
       n_2 & 1 & n^{\prime}_2 \\
       -m_2 & p_2 & m^{\prime}_2
   \end{pmatrix}
     \begin{pmatrix}
       n_2 & 1 & n^{\prime}_2 \\
      0 & 0 & 0
   \end{pmatrix}.  \label{eq:v_dd_basis2}
\end{align}

In the presence of an external static electric field $F$ along $Z$, the rotational states are field-dressed due to the Stark interaction. They are denoted $|\tilde{n},m\rangle$ and correlate at zero field to free-rotor states $|n,m\rangle$. The projection $m$ is conserved and
\begin{equation}\label{eq:dressed}
|\tilde{n},m\rangle = \sum_n c^m_{n \tilde{n}}(F) |n,m\rangle.
\end{equation}
The matrix elements of the single-molecule dipole operator between the field-dressed functions are
\begin{align}\label{eq:dipoles}
    d_{\tilde{n}m,\tilde{n}'m'} (F) &= (-1)^m\mu \sum_{n n^{\prime}} c^m_{n \tilde{n}}  c^{m^{\prime}}_{n^{\prime} \tilde{n}^{\prime}}\sqrt{(2n+1)(2n^{\prime}+1)} \nonumber \\
   & \times
    \begin{pmatrix}
        n & 1 & n^{\prime} \\
        -m & m-m^{\prime} & m^{\prime}
    \end{pmatrix}
    \begin{pmatrix}
        n & 1 & n^{\prime} \\
        0 & 0 & 0
    \end{pmatrix}.
\end{align}
The matrix elements of $\hat{H}_\textrm{dd}$ in the field-dressed rotor pair basis set are
\begin{align}\label{eq:v_dd_simp}
  & \langle \tilde{n}_1, m_1; \tilde{n}_2, m_2 | \hat{H}_\textrm{dd} | \tilde{n}^{\prime}_1, m^{\prime}_1;
\tilde{n}^{\prime}_2, m^{\prime}_2 \rangle \nonumber \\
&= -\frac{\sqrt{30}d_{\tilde{n}_1m_1,\tilde{n}'_1m'_1} d_{\tilde{n}_2m_2,\tilde{n}'_2m'_2}}{4\pi\epsilon_0R^3}
    \begin{pmatrix}
        2 & 1 & 1 \\
        p & -p_1 & -p_2
    \end{pmatrix}C_{2,-p}(\theta, \phi).
\end{align}

For identical molecules, the rotor pair functions are symmetrised with respect to exchange,
\begin{align}
| \tilde{n}_1, m_1; \tilde{n}_2, m_2 \rangle^\pm &= \left[2(1+\delta_{\tilde{n}_1\tilde{n}_2}\delta_{m_1m_2})\right]^{-1/2}
\nonumber\\
& \times \left(| \tilde{n}_1, m_1; \tilde{n}_2, m_2 \rangle \pm | \tilde{n}_2, m_2; \tilde{n}_1, m_1 \rangle\right).
\end{align}
If the two molecules are initially in the same state, as is usual for shielding, only the + linear combination contributes. This is true for either bosons or fermions.

In a full treatment of molecular collisions \cite{Mukherjee:CaF:2023}, we construct basis functions that are products of rotor pair functions $|\tilde{n}_1, m_1; \tilde{n}_2, m_2\rangle^\pm$ and partial-wave functions $|L,M_L\rangle$. Expanding the total Schr\"odinger equation in this basis set gives a set of coupled-channel equations. These may be solved as in ref.\ \cite{Mukherjee:CaF:2023} to produce S-matrices, scattering lengths and cross sections for elastic scattering and 2-body loss. In the present work, however, we bypass coupled-channel calculations and focus instead on effective potentials as described below.

\subsection{Shielding with a static electric field}\label{ssec:dcfield}

\begin{figure}
\begin{center}
	\includegraphics[width=0.4\textwidth]{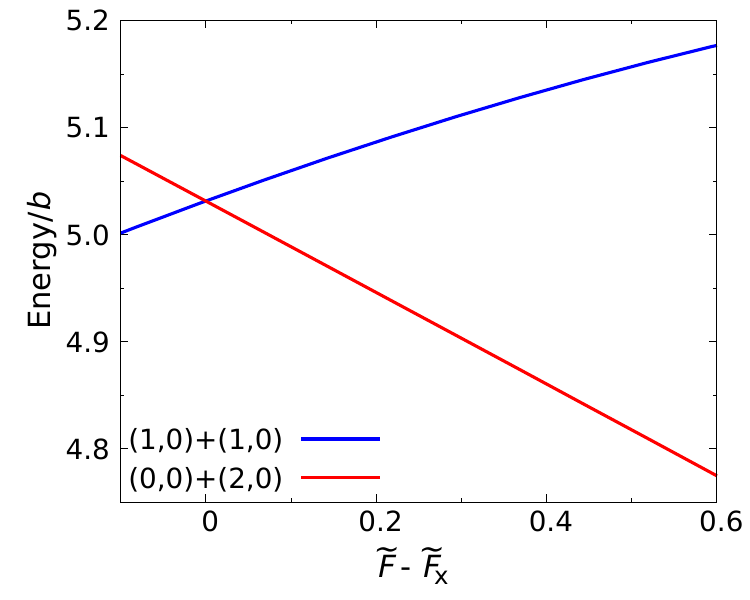}
    \caption{The crossing of pair states (1,0)+(1,0) and (0,0)+(2,0) that gives rise to static-electric field shielding. (1,0)+(1,0) is the initial pair state of the molecules. Energies and fields are shown in dimensionless units; in this form the figure applies to any linear molecule in a $^1\Sigma$ state, with $\tilde{F}_\textrm{X}=3.244$.}
    \label{fig:stark}
\end{center}
\end{figure}

For collisions shielded by a static electric field, it is convenient to represent the field by a dimensionless quantity $\tilde{F} = F\mu/b$~\cite{Gonzalez-Martinez:adim:2017}, where $b$ is the rotational constant of the molecule. Similarly, the molecular energies are expressed in units of $b$.
The pair state $(\tilde{n}_1,m_1)$+$(\tilde{n}_2,m_2)=(1,0)$+(1,0) crosses (0,0)+(2,0) at a critical electric field $\tilde{F}_\textrm{X} = 3.244$, as shown in Fig.\ \ref{fig:stark}. At fields $\tilde{F}>\tilde{F}_\textrm{X}$, (1,0)+(1,0) lies above (0,0)+(2,0), and thus experiences repulsion due to interaction with it via $\hat{H}_\textrm{dd}$. This gives rise to \textit{shielding} for the initial pair state (1,0)+(1,0).

For two dipoles fixed in space, the coupled equations may be cast in dimensionless form \cite{Bohn:BCT:2009}. Here we choose to express lengths in terms of a dipole length $R_3 = (2\mu_\textrm{red}/\hbar^2) (\mu^2/4\pi\epsilon_0)$ and energies in terms of a corresponding energy $E_3 = \hbar^2/(2\mu_\textrm{red} R_3^2)$. However, static-field shielding arises from mixing of rotational states. At a particular value of $\tilde{F}$, \emph{all} the separations between different rotor states are proportional to $b$, or in reduced units to $\tilde{b}=b/E_3$. This includes the separation between (1,0)+(1,0) and (0,0)+(2,0). Because of this, static-field shielding depends crucially on $\tilde{b}$, which ranges from ${\sim}10^6$ for NaLi to ${\sim}10^{11}$ for YO \cite{Gonzalez-Martinez:adim:2017}. Molecules with $\tilde{b}>10^9$ show effective shielding properties at fields $\tilde{F}_\textrm{X}<\tilde{F}<\tilde{F}_\textrm{X}+0.6$. Some molecules with smaller $\tilde{b}$ are still promising for shielding but at fields in a somewhat smaller range~\cite{Gonzalez-Martinez:adim:2017, Mukherjee:alkali:2024}. In the following, we consider effective potentials for both CaF (with $\tilde{b}=4.1\times 10^9$) and RbCs (with $\tilde{b}=2.6\times 10^8$) to illustrate the differences that arise.

\begin{figure}
\begin{center}
	\includegraphics[width=0.42\textwidth]{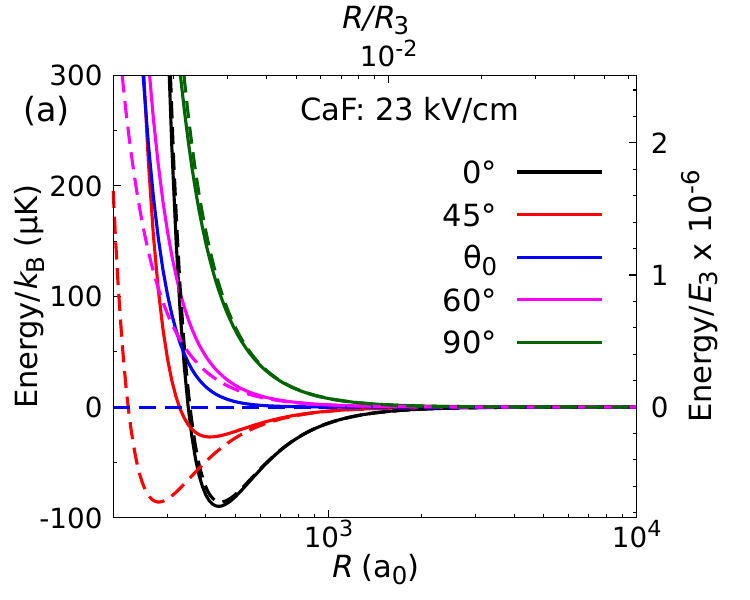}
    \includegraphics[width=0.42\textwidth]{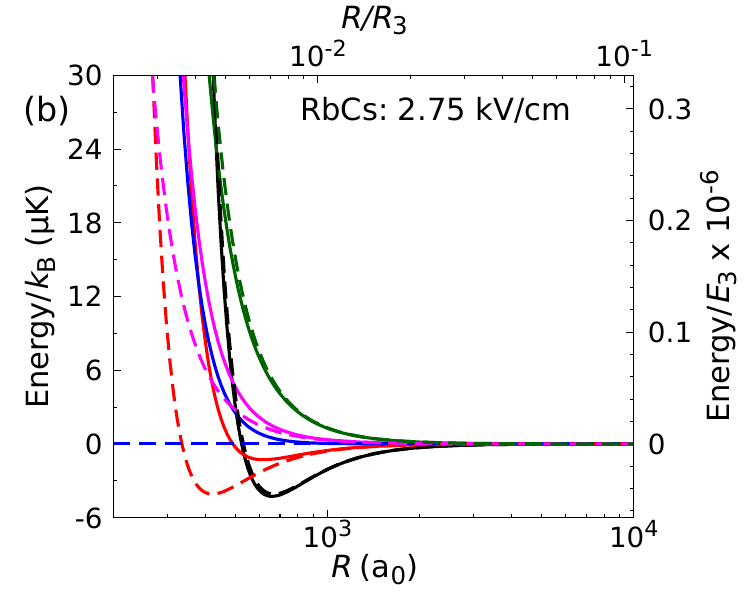}
    \caption{The adiabats that correspond to the initial pair state (1,0)+(1,0) for (a) CaF and (b) RbCs at electric field $\tilde{F}=3.46$. Solid lines are obtained from full matrix diagonalisation whereas dashed lines are obtained by diagonalising the matrix of Eq.\ \ref{eq:hdd_npair2}.}
    \label{fig:full_adiabats}
\end{center}
\end{figure}

\begin{figure}
\begin{center}
	\includegraphics[width=0.42\textwidth]{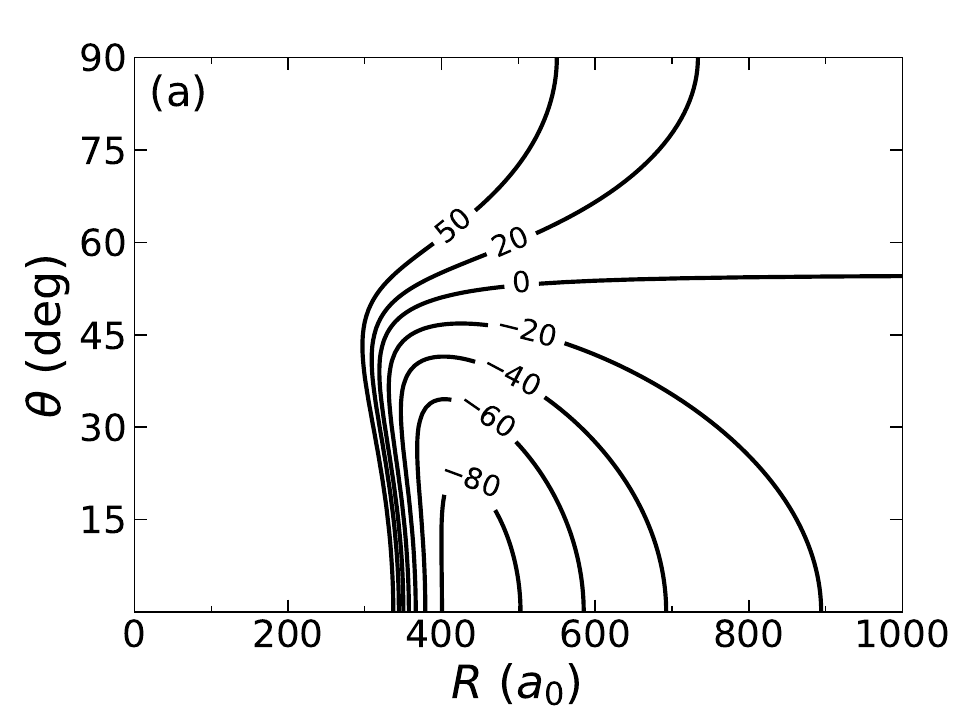}
    \includegraphics[width=0.42\textwidth]{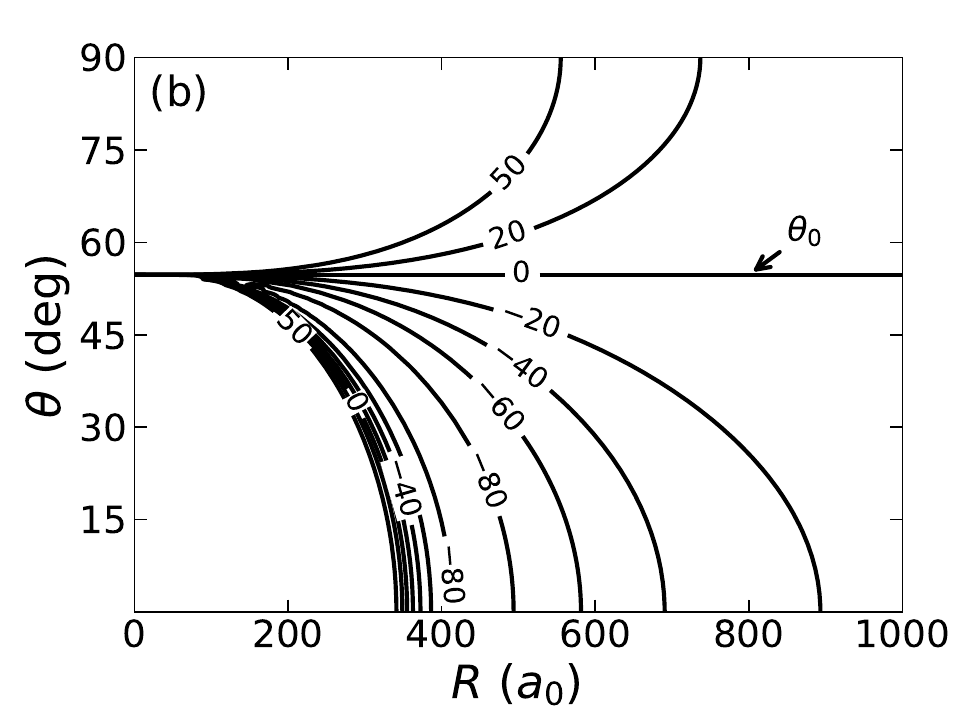}
    \caption{Contour plots of adiabats obtained from (a) full matrix diagonalisation, and (b) diagonalising the matrix of Eq.\ \ref{eq:hdd_npair2} for CaF at $F=23$ kV/cm. Contours are labeled in $\mu$K and shown only up to 50 $\mu$K.}
    \label{fig:contours_LQ}
\end{center}
\end{figure}

We begin by constructing a basis set of field-dressed rotor pair functions $|\tilde{n}_1, m_1; \tilde{n}_2, m_2 \rangle^+$, including functions up to $n_\textrm{max}=\tilde{n}_\textrm{max}=5$ for each molecule. This gives good numerical convergence. Diagonalising the Hamiltonian matrix formed from $\hat{h}_1+\hat{h}_2+\hat{H}_\textrm{dd}$ in this basis set produces adiabats (effective potentials) that are functions of $R$ and $\theta$. There is no dependence on $\phi$ because the two-molecule system is cylindrically symmetric in the presence of an electric field. Adiabats corresponding to the initial pair state (1,0)+(1,0) for CaF and RbCs are shown as solid lines in Fig.\ \ref{fig:full_adiabats} for a field where shielding is effective. The results for CaF are shown as a contour plot in Fig.\ \ref{fig:contours_LQ}(a). The effective potentials are repulsive inside $R\sim 300\ a_0$ at all angles, but are attractive at long range for angles $\theta<\theta_0$, where $\theta_0=\cos^{-1}(1/\sqrt{3})\approx 55^\circ$ is the semi-tetrahedral angle where the first-order dipole-dipole interaction vanishes.

The effective potentials are qualitatively similar for CaF and RbCs, but the positions and depths of the potential wells are quite different in reduced units. In these units the first-order dipole-dipole part of the potential is the same for the two species, but the shielding repulsion is approximately proportional to $1/\tilde{b}$. As a result, the potential wells for CaF are much deeper and occur at much shorter range than those for RbCs in reduced units.

\section{Analytic effective potentials}
\label{sec:derivation}

Our goal is to obtain a simple approximate form of the effective potential for the initial pair state (1,0)+(1,0), avoiding diagonalisation of substantial matrices. We first consider a $2\times2$ Hamiltonian matrix that involves only the pair states (1,0)+(1,0) and (0,0)+(2,0), here denoted 1 and 2, respectively. These are the states that produce the main shielding effect at most angles. Using Eq.\ \ref{eq:v_dd_simp}, the matrix elements may be written
\begin{subequations}\label{eq:hdd_npair2}
\begin{align}
  H_{11} &= -\frac{2(d_{10, 10})^2}{4\pi\epsilon_0 R^3} P_2(\cos\theta), \\
  H_{12} &= -\frac{2\sqrt{2}d_{00, 10}d_{10, 20}}{4\pi\epsilon_0 R^3} P_2(\cos\theta), \\
  H_{22} &= -\frac{2\Big(d_{00, 00}d_{20, 20}+(d_{00, 20})^2\Big)}{4\pi\epsilon_0 R^3}P_2(\cos\theta) - \Delta,
\end{align}
\end{subequations}
where $P_2(\cos\theta)= \frac{1}{2}(3\cos^2 \theta -1)$. The matrix elements $d_{\tilde{n}m, \tilde{n}'m'}$ and the energy separation $\Delta$ between states 1 and 2 are functions of field.

Lassabli\`ere and Qu\'em\'ener \cite{Lassabliere:2022} obtained effective potentials by diagonalising the $2\times 2$ matrix of Eq.\ \ref{eq:hdd_npair2}. The resulting adiabats are shown as dashed lines for CaF and RbCs in Fig.\ \ref{fig:full_adiabats}. Figure \ref{fig:contours_LQ}(b) shows the results for CaF as a contour plot. It may be seen that the potentials of ref.\ \cite{Lassabliere:2022} are accurate at $\theta=0$ and $90^\circ$, but show no attraction or repulsion at the semi-tetrahedral angle $\theta_0$, where $P_2(\cos\theta)=0$. They produce attractive wells that are substantially too deep and at much too short range at angles just below $\theta=\theta_0$. These properties would be likely to produce unphysical artefacts in simulations of many-body properties.

To overcome this limitation, it is necessary to include additional pair states that provide repulsion around $\theta=\theta_0$. These are energetically further away and we take them into account through either second-order perturbation theory or a Van Vleck transformation.

In the present work, we first describe a minimal model for the effective potential in Sec.\ \ref{ssec:perturbative}, using second-order perturbation theory in a limited basis set. Subsequently, in Sec.\ \ref{ssec:coupled}, we describe a more accurate treatment, based on Van Vleck perturbation theory in a larger basis set, that is only slightly more complicated to evaluate and gives accurate results.

\subsection{Perturbative model}
\label{ssec:perturbative}

We obtain an effective potential using a perturbative model $V^\textrm{pert}_\textrm{eff}$ by considering the initial pair state (1,0)+(1,0) only and treating the most important additional pair states using second-order perturbation theory. We write this effective potential as
\begin{align}\label{eq:perturb}
  V^\textrm{pert}_\textrm{eff}(R,\theta) &= H_{11} (R,\theta) + \frac{|H_{12}(R,\theta)|^2}{H_{11}-H_{22}}
  \nonumber \\
  &+ V_\textrm{rest}(R,\theta).
\end{align}
The second term represents the repulsion due to the lower-lying pair state (0,0)+(2,0), while $V_\textrm{rest}(R,\theta)$ takes account of the interaction due to other pairs. In this model, both of these are treated perturbatively.

\begin{figure}
\begin{center}
	\includegraphics[width=0.42\textwidth]{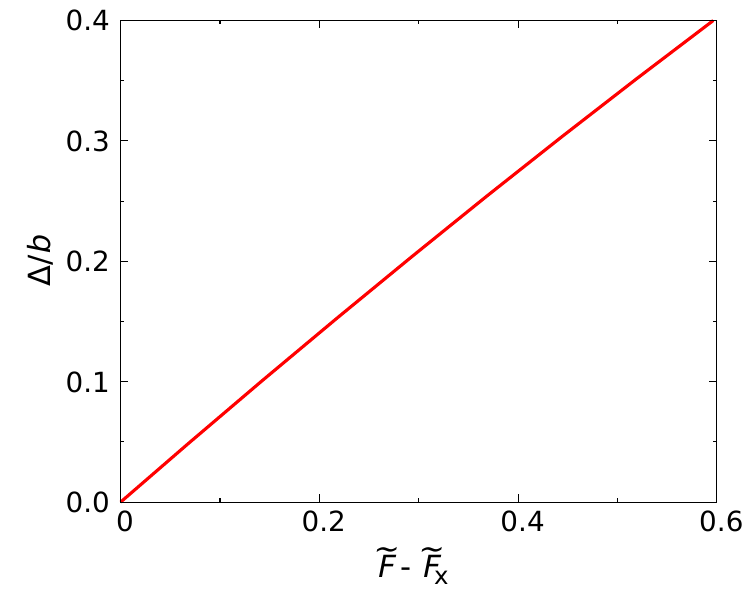}
    \caption{The energy separation between pair states (1,0)+(1,0) and (0,0)+(2,0). The field range is chosen where shielding is effective.}%
    \label{fig:delta}
\end{center}
\end{figure}

We formulate the effective potential for fields $\tilde{F}_\textrm{X}<\tilde{F} <\tilde{F}_\textrm{X}+0.6$, where shielding is effective for molecules such as CaF. For RbCs, the range of effective shielding is somewhat smaller, extending up to $\tilde{F}_\textrm{X}+0.4$.

At most fields, the matrix elements of $\hat{H}_\textrm{dd}$ are much smaller than the asymptotic separation $\Delta$ between states 1 and 2. We therefore make the approximation
\begin{align}\label{eq:delta}
  H_{11}-H_{22} &\approx \Delta.
\end{align}
Figure \ref{fig:delta} shows that $\Delta$ is very close to linear in $\tilde{F}-\tilde{F}_\textrm{X}$ in the field range where shielding is effective. We therefore write
\begin{align}\label{eq:deltafit}
  \Delta/b &\approx \delta (\tilde{F}-\tilde{F}_\textrm{X}),
\end{align}
with slope $\delta=0.684$.

\begin{figure}
\begin{center}
	\includegraphics[width=0.42\textwidth]{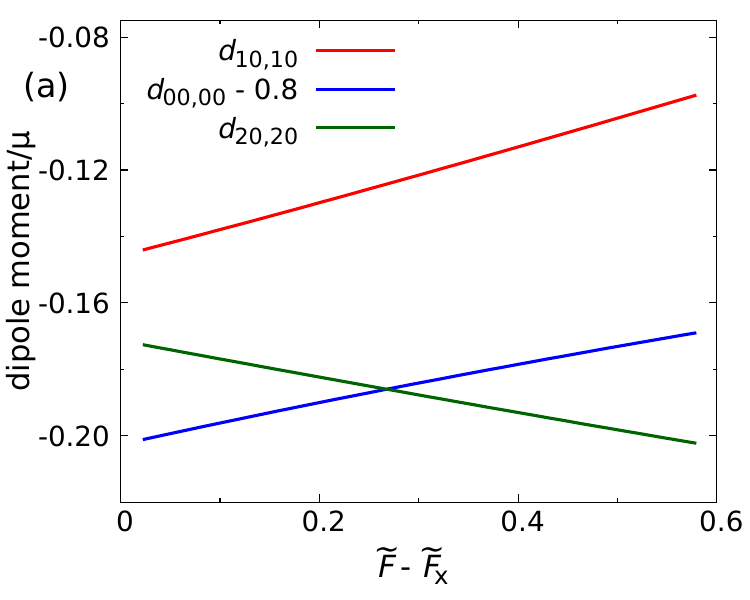}
    \includegraphics[width=0.42\textwidth]{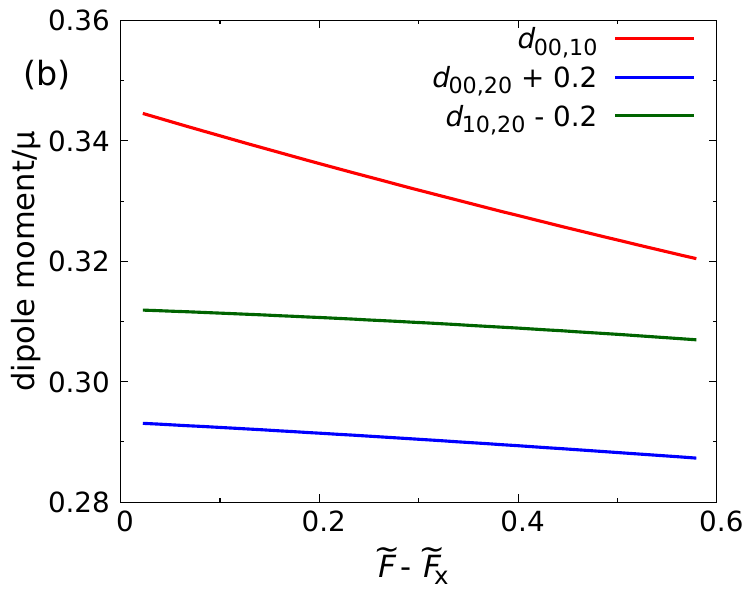}
    \caption{The single-molecule dipole matrix elements that appear in the dipole-dipole matrix elements of Eq.\ \ref{eq:hdd_npair2}. The blue curves in both panels and the green curve in (b) are shifted to fit within the scales of the figures.}%
    \label{fig:dipoles}
\end{center}
\end{figure}

The dipole moments for the monomer states $(\tilde{n},m_n)=(0,0)$, (1,0) and (2,0), and the transition dipole moments appearing in the matrix elements of Eqs.\ \ref{eq:hdd_npair2} are shown in Fig.\ \ref{fig:dipoles}. For simplicity, we denote the dipole matrix elements $d_{i0,j0}$ as $d_{ij}$. As seen in Fig.\ \ref{fig:dipoles}, these are approximately linear functions of $F$ in the region of interest, so we approximate each of them as
\begin{align}\label{eq:dipfit}
  d_{ij}/\mu \approx d^{(0)}_{ij} + d^{(1)}_{ij}(\tilde{F}-\tilde{F}_\textrm{X}),
\end{align}
with the coefficients given in Table \ref{tab:params}.

\begin{table}[tbp]
\caption{The parameters defining the effective potentials obtained from the perturbative model of Eq.\ \ref{eq:veff}, and the coupled model of Eq.\ \ref{eq:Veff_coup}.
\label{tab:params}} \centering
\begin{ruledtabular}
\begin{tabular}{cc}
Parameter & Value \\
\hline
$\delta$ & \,\, 0.684 \\
$d^{(0)}_{00}$ & \,\, 0.598 \\
$d^{(1)}_{00}$ & \,\,  0.058 \\
$d^{(0)}_{11}$ & $-0.146$ \\
$d^{(1)}_{11}$ & \,\,  0.084 \\
$d^{(0)}_{22}$ & $-0.172$ \\
$d^{(1)}_{22}$ & $-0.053$ \\
$d^{(0)}_{01}$ & \,\, 0.345 \\
$d^{(1)}_{01}$ & $-0.043$ \\
$d^{(0)}_{02}$ & \,\, 0.093 \\
$d^{(1)}_{02}$ & $-0.010$ \\
$d^{(0)}_{12}$ & \,\, 0.512 \\
$d^{(1)}_{12}$ & $-0.009$ \\
$\alpha^{(0)}$ & \,\, 0.428 \\
$\alpha^{(1)}$ &  $-1.417$ \\
$\alpha^{(2)}$ & \,\, 1.315
\end{tabular}
\end{ruledtabular}
\end{table}

\begin{figure}[tbp]
\begin{center}
	\includegraphics[width=0.42\textwidth]{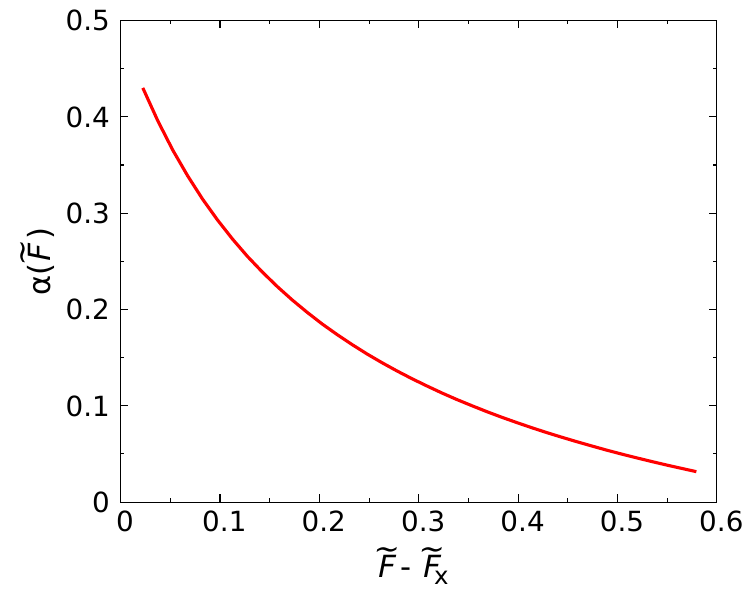}
    \caption{Functional form of $\alpha(\tilde{F})$.}%
    \label{fig:alpha}
\end{center}
\end{figure}

\begin{figure*}[tbp]
\begin{center}
    \includegraphics[width=0.3\textwidth]{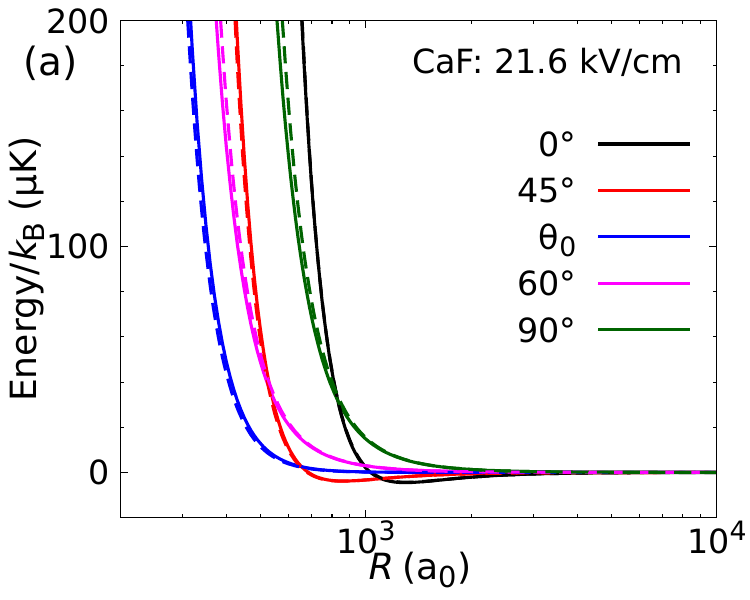}
    \includegraphics[width=0.3\textwidth]{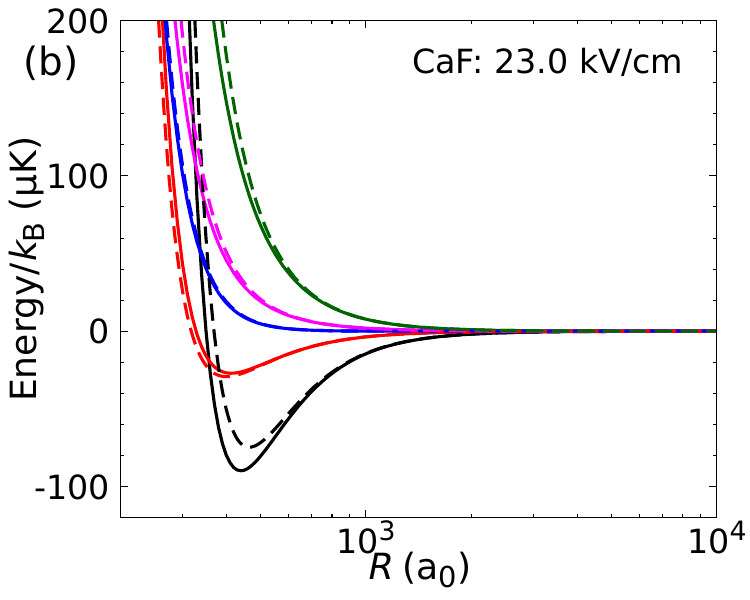}
    \includegraphics[width=0.3\textwidth]{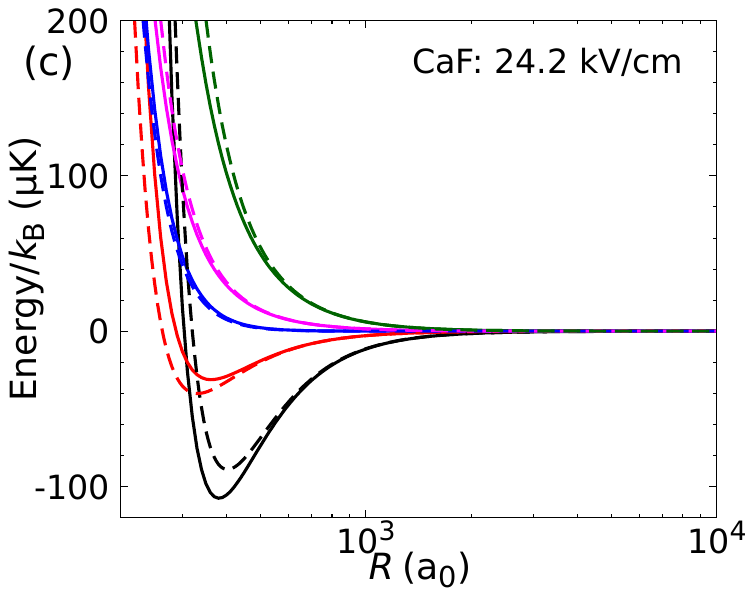}
    \includegraphics[width=0.3\textwidth]{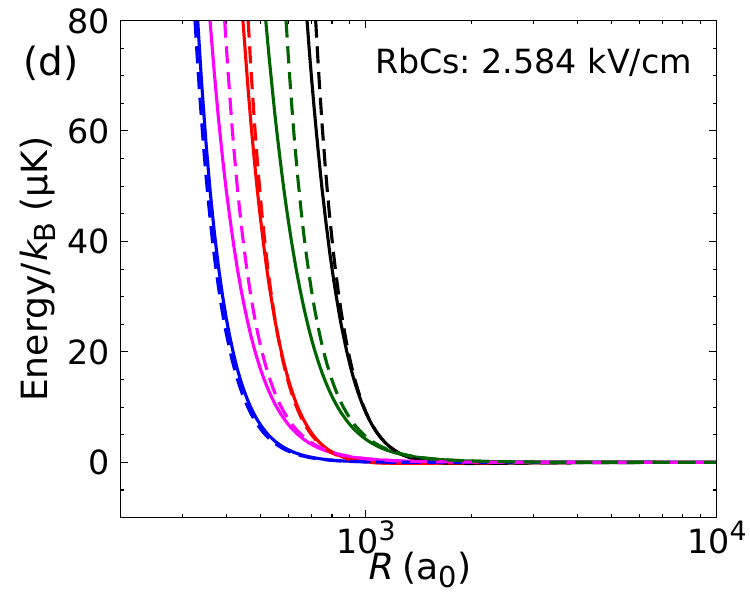}
    \includegraphics[width=0.3\textwidth]{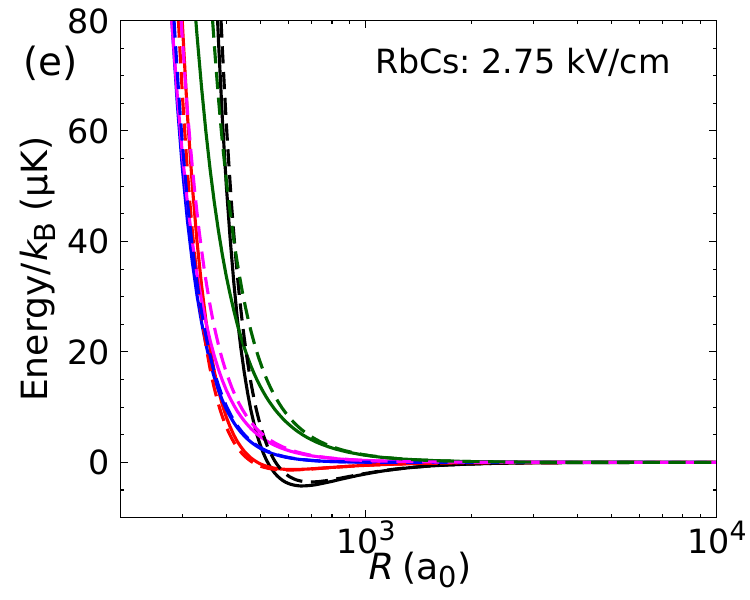}
    \includegraphics[width=0.3\textwidth]{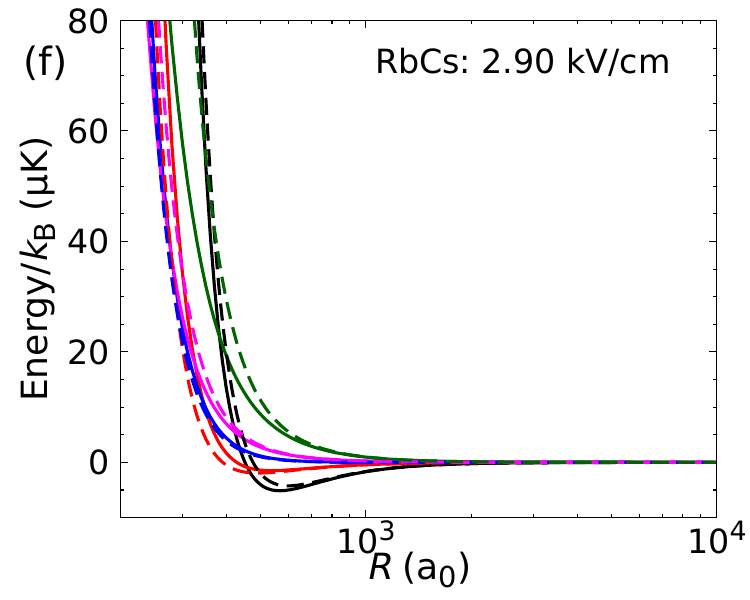}
    \caption{Comparison of adiabats for CaF and RbCs in the initial pair state (1,0)+(1,0), obtained from the perturbative model of Eq.\ \ref{eq:veff} (dashed lines) and from full matrix diagonalisation (solid lines). The values of $\tilde{F}$ correspond to (a) and (d) 3.25; (b) and (e) 3.46; and (c) and (e) 3.65.}
    \label{fig:veff_pert_compare}
\end{center}
\end{figure*}

The contribution $V_\textrm{rest}(R,\theta)$ from other pair states $k$ is represented as
\begin{align}\label{eq:hpert}
 V_\textrm{rest}(R,\theta) &= \sum_k \frac{|H_{1k}|^2}{E_1-E_k}.
\end{align}
This is smaller than the first two terms of Eq.\ \ref{eq:perturb} at most $\theta$, except near $\theta=\theta_0$, where both $H_{11}$ and $H_{12}$ vanish. In the perturbative model, for simplicity, we therefore evaluate $V_\textrm{rest}$ only at $\theta=\theta_0$ and use the result at all $\theta$. At $\theta=\theta_0$, pair states with $m_1'+m_2'=0$ make no contribution and $V_\textrm{rest}(R)$ is dominated by a repulsive interaction due to (0,0)+(2,$\pm$1), which is the pair state that lies closest to (and below) (0,0)+(2,0) at these fields. All the state separations are proportional to the rotational constant $b$ except for a weak dependence on $R$. With this approximation,
\begin{equation}\label{eq:H_11VV}
 V_\textrm{rest}(R) \equiv V_\textrm{rest}(R,\theta_0) = \frac{\mu^4}{(4\pi\epsilon_0)^2R^6}\frac{\alpha}{b}.
\end{equation}
The dimensionless function $\alpha(\tilde{F})$ is shown in Fig.\ \ref{fig:alpha}. We approximate it as a quadratic function of $F$,
\begin{align}\label{eq:alpha}
  \alpha(\tilde{F}) &\approx \alpha^{(0)} + \alpha^{(1)}(\tilde{F}-\tilde{F}_\textrm{X}) + \alpha^{(2)}(\tilde{F}-\tilde{F}_\textrm{X})^2,
\end{align}
with the coefficients given in Table \ref{tab:params}.

Finally, we write the expression for the effective potential from Eq.\ \ref{eq:perturb} as
\begin{align}\label{eq:veff}
  V^\textrm{pert}_\textrm{eff} (R,\theta) = \frac{C_3(\theta)}{R^3}+\frac{C_6 (\theta)}{R^6},
\end{align}
where the coefficients are
\begin{align}\label{eq:coeffc3}
  C_3(\theta) &= -\frac{2d_{11}^2}{4\pi\epsilon_0}P_2(\cos\theta),
\end{align}
and
\begin{equation}\label{eq:coeffc6}
  C_6(\theta) = \frac{1}{(4\pi\epsilon_0)^2}\Bigg( \frac{8[d_{01} d_{12} P_2(\cos\theta)]^2}{\Delta}  + \frac{\alpha\mu^4}{b} \Bigg),
\end{equation}
with $\Delta(\tilde{F})$, $d_{ij}(\tilde{F})$ and $\alpha(\tilde{F})$ from Eqs.\ \ref{eq:deltafit}, \ref{eq:dipfit} and \ref{eq:alpha}, respectively.

\begin{figure}
\begin{center}
	\includegraphics[width=0.45\textwidth]{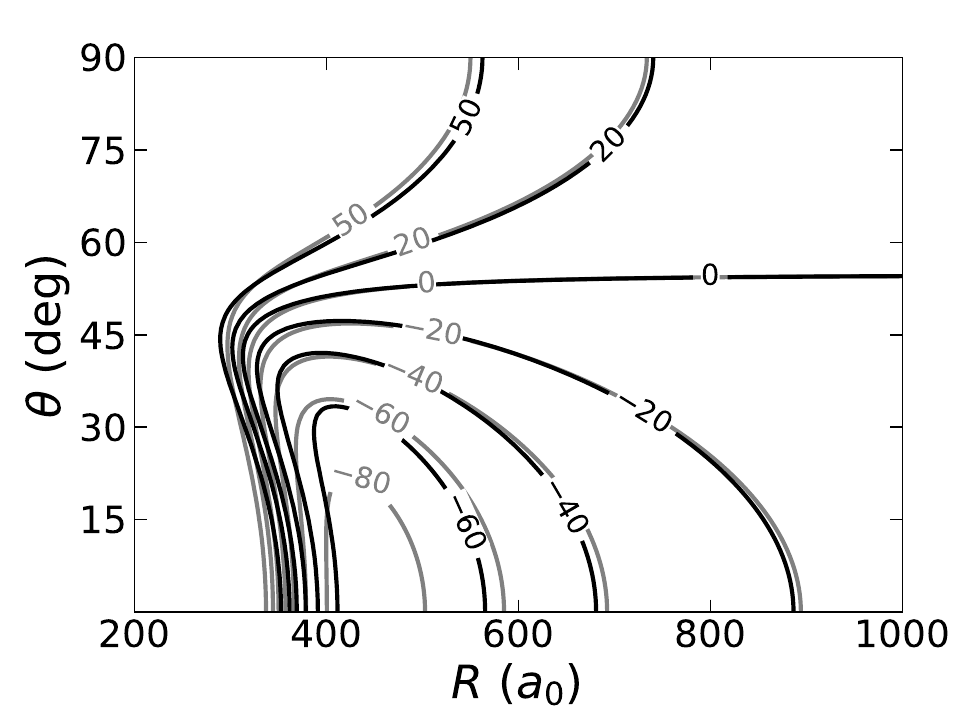}
    \caption{Contour plot comparing $V^\textrm{pert}_\textrm{eff}$ (black lines) with adiabats obtained from full matrix diagonalisation (grey lines) for CaF at $F=23$ kV/cm. Contours are labeled in $\mu$K and shown only up to 50 $\mu$K. The potential minimum for $V^\textrm{pert}_\textrm{eff}$ is at $-75\ \mu$K, so the contour for $-80\ \mu$K does not appear.}
    \label{fig:contour_pert}
\end{center}
\end{figure}

Figure \ref{fig:veff_pert_compare} compares the perturbative model with adiabats obtained from full matrix diagonalisation. Results are shown for CaF and RbCs in the initial pair state (1,0)+(1,0) at three different fields where shielding is effective. Figure \ref{fig:contour_pert} shows the comparison for CaF at 23 kV/cm as contour plots.

The perturbative model reproduces the qualitative features of the adiabats well. It corrects the main deficiency of Fig.\ \ref{fig:contours_LQ}(b) without significant extra computational expense. However, it produces some quantitative errors that may be significant in precise work. Its main deficiency is that it approximates the effect of pair states with $m_1'+m_2'\ne 0$ as isotropic, as in Eq.\ \ref{eq:H_11VV}. Furthermore, it neglects the effects of pair states other than (0,0)+(2,0) with $m_1'+m_2'=0$ entirely, because Eq.\  \ref{eq:H_11VV} is evaluated at $\theta_0$, where $P_2(\cos\theta)=0$. The approximation becomes more serious at larger $\tilde{F}$, where the effects of (0,0)+(2,0) are not so strongly dominant. As seen in Fig.\  \ref{fig:veff_pert_compare}, it can cause underestimates of up to 20\% in well depths at higher fields. It also starts to break down at small distances when the off-diagonal matrix element $H_{12}$ in Eq.\ \ref{eq:hdd_npair2} is comparable to the energy separation $\Delta$ between pair states 1 and 2. However, this occurs only for fields very close to $F_\textrm{X}$ and is most significant for molecules with small values of $\tilde{b}$. It may be seen in the lowest-field panels in Fig.\ \ref{fig:veff_pert_compare} (for $\tilde{F}=3.25$); the repulsion at angles different from $\theta_0$ is overestimated for RbCs at 2.584 kV/cm but is reasonably accurate for CaF at 21.6 kV/cm.

\subsection{Fully coupled model}\label{ssec:coupled}

The perturbative model described in section \ref{ssec:perturbative} gives qualitatively correct results, but with some deviations from full diagonalisation. To improve on it, we need to consider the effects of additional pair states. We do this using Van Vleck perturbation theory~\cite{VanVleck:1929, Kemble:1937}. In this approach, the full set of basis functions is partitioned into two classes, denoted class 1 (labels $i$, $j$, $\ldots$) and class 2 (labels $\alpha$, $\beta$, $\ldots$). The most important functions are in class 1 and no function in class 2 is close in energy to any function in class 1. The functions in class 1 are included explicitly in matrix diagonalisations, while those in class 2 are included perturbatively. This produces matrix elements between different functions in class 1 of the form
\begin{align}\label{eq:VV}
\langle i &| \hat{H}_\textrm{dd,VV} | j \rangle \nonumber\\
&= \sum_\alpha \frac{1}{2} \left[
\frac{\langle i | \hat{H}_\textrm{dd} | \alpha \rangle
\langle \alpha | \hat{H}_\textrm{dd} | j \rangle} {(E_i-E_\alpha)}
+ \frac{\langle i | \hat{H}_\textrm{dd} | \alpha \rangle
\langle \alpha | \hat{H}_\textrm{dd} | j \rangle} {(E_j-E_\alpha)}
\right].
\end{align}
Since $\hat{H}_\textrm{dd}$ is proportional to $R^{-3}$, $\hat{H}_\textrm{dd,VV}$ is proportional to $R^{-6}$.

\begin{table}[tbp]
\caption{The fitted values of the parameters in Eqs.\ \ref{eq:pqr}.
\label{tab:VVparams}} \centering
\begin{ruledtabular}
\begin{tabular}{cc}
Parameter & Value \\
\hline
$P_{11}^{(0)}$ & $-1.76 \times 10^{-2}$ \\
$P_{11}^{(1)}$ & $-2.90 \times 10^{-3}$ \\
$P_{22}^{(0)}$ & \,\,\, $1.46 \times 10^{-2}$ \\
$P_{22}^{(1)}$ & $-2.11 \times 10^{-3}$ \\
$P_{12}^{(0)}$ & $-1.14 \times 10^{-2}$ \\
$P_{12}^{(1)}$ & \,\,\, $2.76 \times 10^{-3}$ \\
$Q_{11}^{(0)}$ & \,\, $2.07$ \\
$Q_{11}^{(1)}$ & $-6.31$ \\
$Q_{11}^{(2)}$ & \,\, $5.92$ \\
$Q_{22}^{(0)}$ & \,\,\, $1.95 \times 10^{-1}$ \\
$Q_{22}^{(1)}$ & \,\,\, $1.44 \times 10^{-1}$ \\
$Q_{22}^{(2)}$ & \,\,\, $1.90 \times 10^{-2}$ \\
$Q_{12}^{(0)}$ & $-4.45 \times 10^{-1}$ \\
$Q_{12}^{(1)}$ & \,\,\, $5.22 \times 10^{-1}$ \\
$Q_{12}^{(2)}$ & $-5.19 \times 10^{-1}$ \\
$R_{11}^{(0)}$ & $-7.36 \times 10^{-2}$ \\
$R_{11}^{(1)}$ & $-3.36 \times 10^{-2}$ \\
$R_{22}^{(0)}$ & \,\,\, $2.08 \times 10^{-2}$ \\
$R_{22}^{(1)}$ & $-1.90 \times 10^{-2}$ \\
$R_{12}^{(0)}$ & $-4.44 \times 10^{-2}$ \\
$R_{12}^{(1)}$ & $-3.08 \times 10^{-3}$
\end{tabular}
\end{ruledtabular}
\end{table}

In the fully coupled model, we consider the pair functions (1,0)+(1,0) and (0,0)+(2,0) in class 1, and all remaining rotor pairs in class 2. The matrix elements involving states 1 and 2, without the Van Vleck contributions, are given in Eq.\ \ref{eq:hdd_npair2}. We model the matrix elements arising from the Van Vleck transformation as
\begin{align}\label{eq:VVmat}
 H_{ij,\textrm{VV}} (R,\theta,\tilde{F}) &= \frac{\mu^4}{(4\pi\epsilon_0)^2R^6} \frac{1}{b} \bigg[ P_{ij}(\tilde{F})(1-3\cos^2\theta)^2 \nonumber \\
 &+ Q_{ij}(\tilde{F})\sin^2\theta\cos^2\theta + R_{ij}(\tilde{F})\sin^4\theta \bigg],
\end{align}
with $(i,j)=(1,1),(2,2)$ and $(1,2)$.
The first term in square brackets represents second-order matrix elements of $H_\textrm{dd}$ between (1,0)+(1,0) and pair states in class 2 with $|m_1'+m_2'|=0$. These are proportional to the square of the spherical harmonics $|C_{2,0}(\theta,\phi)|^2$. Similarly, the second and third terms represent second-order interactions with pair states in class 2 with $|m_1'+m_2'|=1$ and 2, respectively, and are proportional to $|C_{2,\pm1}(\theta,\phi)|^2$ and $|C_{2,\pm2}(\theta,\phi)|^2$. The coefficients $P(\tilde{F})$, $Q(\tilde{F})$ and $R(\tilde{F})$ are dimensionless functions of field, and are approximated as
\begin{subequations}\label{eq:pqr}
\begin{align}
  P_{ij}(\tilde{F}) &\approx P^{(0)}_{ij} + P^{(1)}_{ij}(\tilde{F}-\tilde{F}_\textrm{X}); \\
  Q_{ij}(\tilde{F}) &\approx Q^{(0)}_{ij} + Q^{(1)}_{ij}(\tilde{F}-\tilde{F}_\textrm{X}) + Q^{(2)}_{ij}(\tilde{F}-\tilde{F}_\textrm{X})^2; \\
  R_{ij}(\tilde{F}) &\approx R^{(0)}_{ij} + R^{(1)}_{ij}(\tilde{F}-\tilde{F}_\textrm{X})
\end{align}
\end{subequations}
with the coefficients given in Table \ref{tab:VVparams}.

\begin{figure*}[tbp]
\begin{center}
    \includegraphics[width=0.3\textwidth]{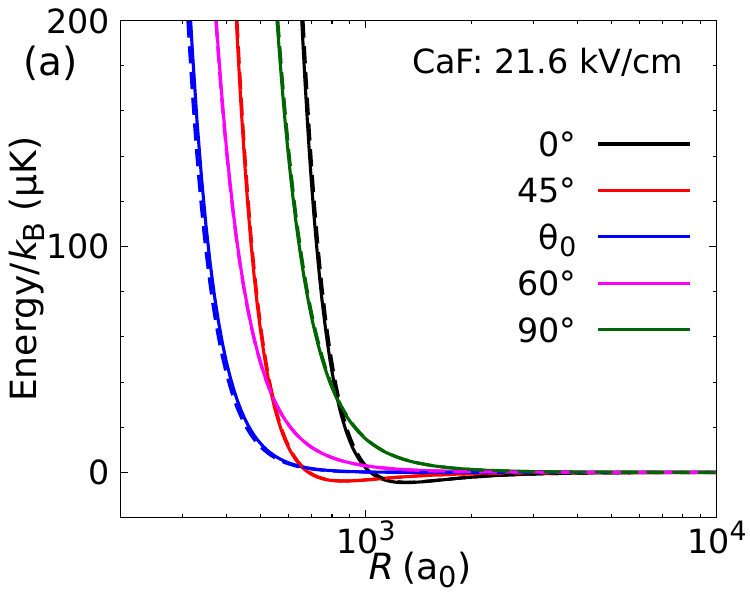}
    \includegraphics[width=0.3\textwidth]{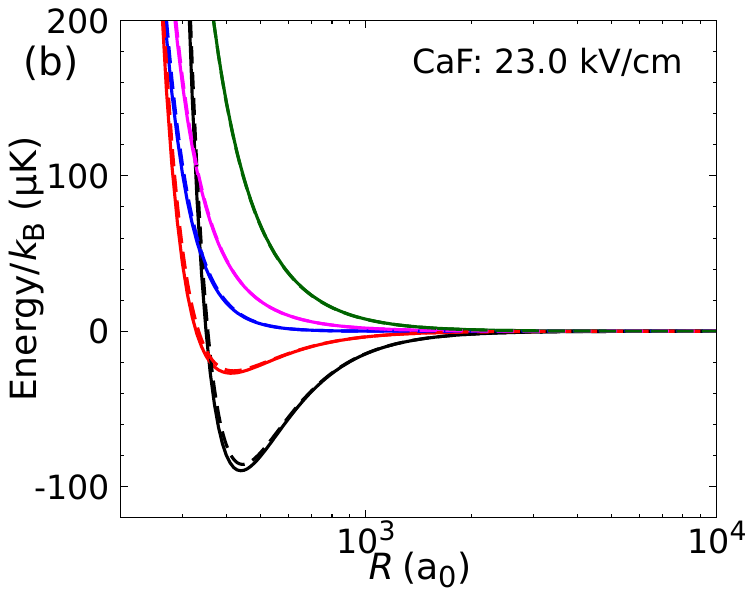}
    \includegraphics[width=0.3\textwidth]{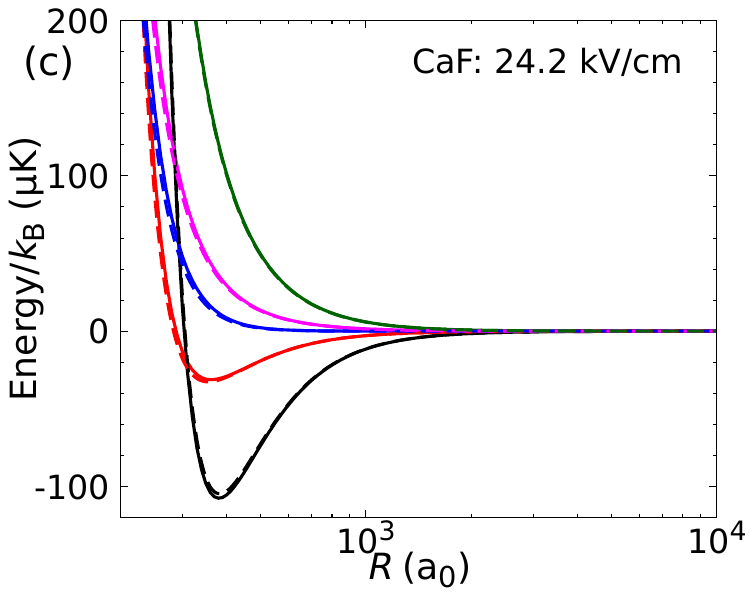}
	\includegraphics[width=0.3\textwidth]{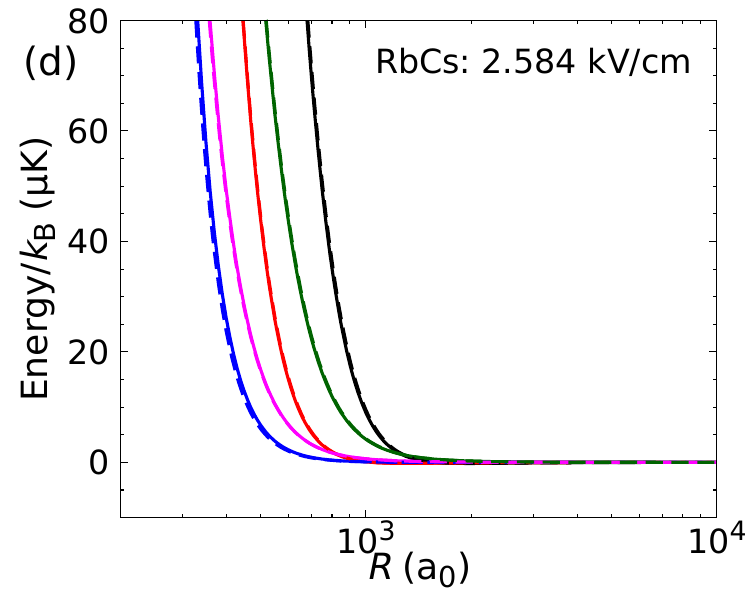}
    \includegraphics[width=0.3\textwidth]{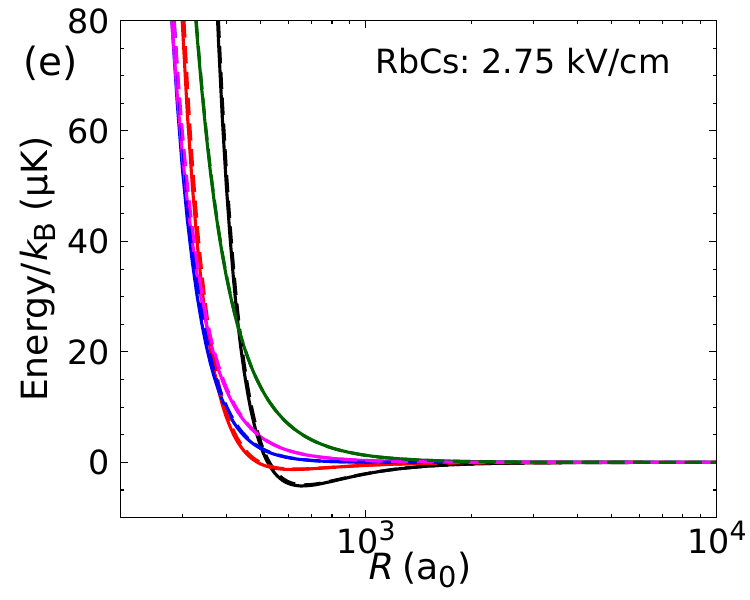}
    \includegraphics[width=0.3\textwidth]{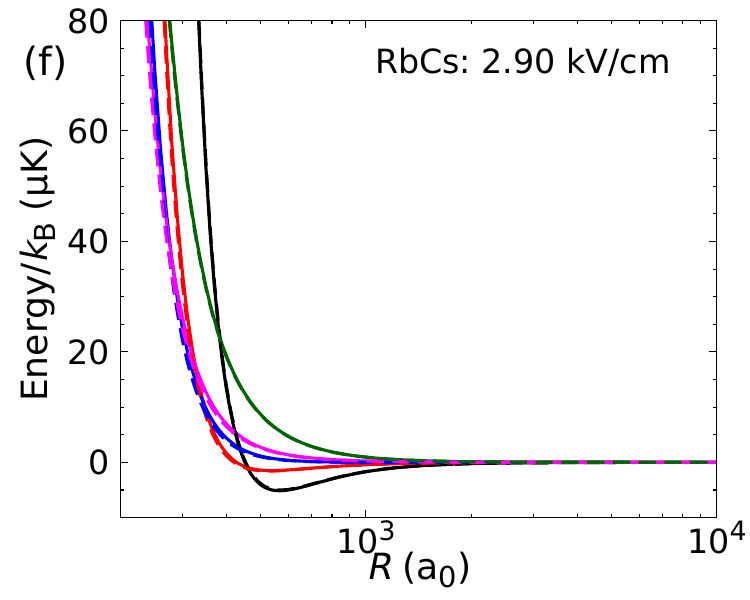}
    \caption{Comparison of adiabats for CaF and RbCs in the initial pair state (1,0)+(1,0), obtained from the fully coupled model of Eq.\ \ref{eq:Veff_coup} (dashed lines) and from full matrix diagonalisation (solid lines). The values of $\tilde{F}$ are the same as in Fig.\ \ref{fig:veff_pert_compare}.}%
    \label{fig:veff_VV_compare}
\end{center}
\end{figure*}

\begin{figure}
\begin{center}
	\includegraphics[width=0.45\textwidth]{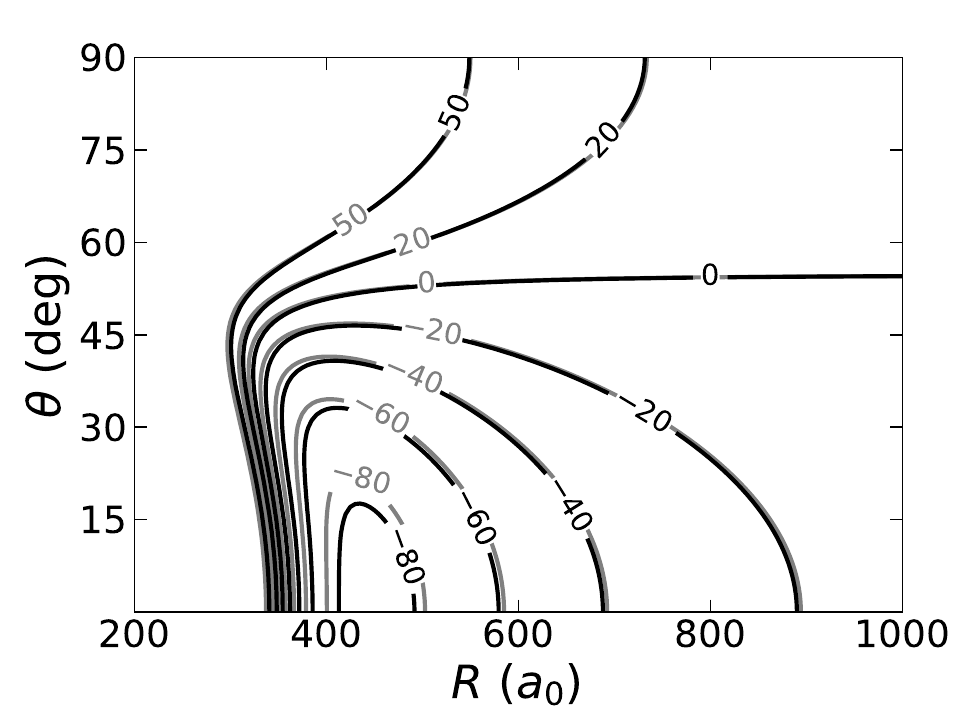}
    \caption{Contour plot comparing $V^\textrm{fc}_\textrm{eff}$ (black lines) with adiabats obtained from full matrix diagonalisation (grey lines) for CaF at $F=23$ kV/cm. Contours are labeled in $\mu$K and shown only up to 50 $\mu$K.}
    \label{fig:contour_fc}
\end{center}
\end{figure}

We supplement the matrix elements $H_{ij}$ of Eq.\ \ref{eq:hdd_npair2} with the Van Vleck matrix elements of Eq.\ \ref{eq:VVmat} to form a $2 \times 2$ matrix $M$ with elements
\begin{equation}\label{eq:Mmat}
  M_{ij} = H_{ij} + H_{ij,\textrm{VV}}.
\end{equation}
The effective potential for (1,0)+(1,0) in the fully coupled model is the higher of the two eigenvalues of $M$,
\begin{align}\label{eq:Veff_coup}
  V^\textrm{fc}_\textrm{eff} (R, \theta) = \frac{1}{2}(M_{11}+M_{22})
  &+ \frac{1}{2}\sqrt{ (M_{11}-M_{22})^2 + 4M_{12}^2 },
\end{align}
where all the matrix elements are functions of $R$, $\theta$ and $\tilde{F}$.

The fully coupled model does not involve any approximations other than the linearisation of the field-dependence of the matrix elements and the Van Vleck treatment itself. Figure \ref{fig:veff_VV_compare} compares $V^\textrm{fc}_\textrm{eff}$ with the full adiabats for CaF and RbCs, while Fig.\ \ref{fig:contour_fc} shows the results for CaF at 23 kV/cm as contour plots. The agreement is excellent, and of similar quality for CaF and RbCs.

\section{Contrast with microwave shielding}

\begin{figure}
\begin{center}
	\includegraphics[width=0.45\textwidth]{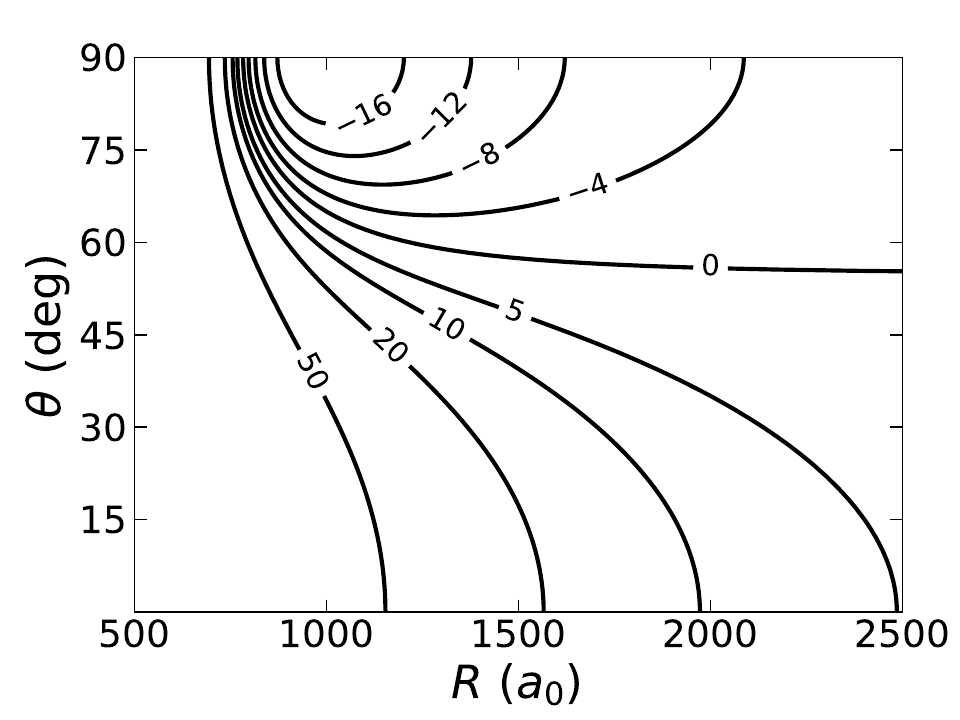}
    \caption{Contour plot of effective potential for shielding of CaF with circularly polarised microwaves at Rabi frequency $\Omega=30$ MHz and detuning $\Delta=0$, using the functional form of ref.\ \cite{Deng:microwave:2023}. Contours are labeled in $\mu$K and shown only up to 50 $\mu$K.}
    \label{fig:contour_MW}
\end{center}
\end{figure}

\begin{figure}
\begin{center}
	\includegraphics[width=0.45\textwidth]{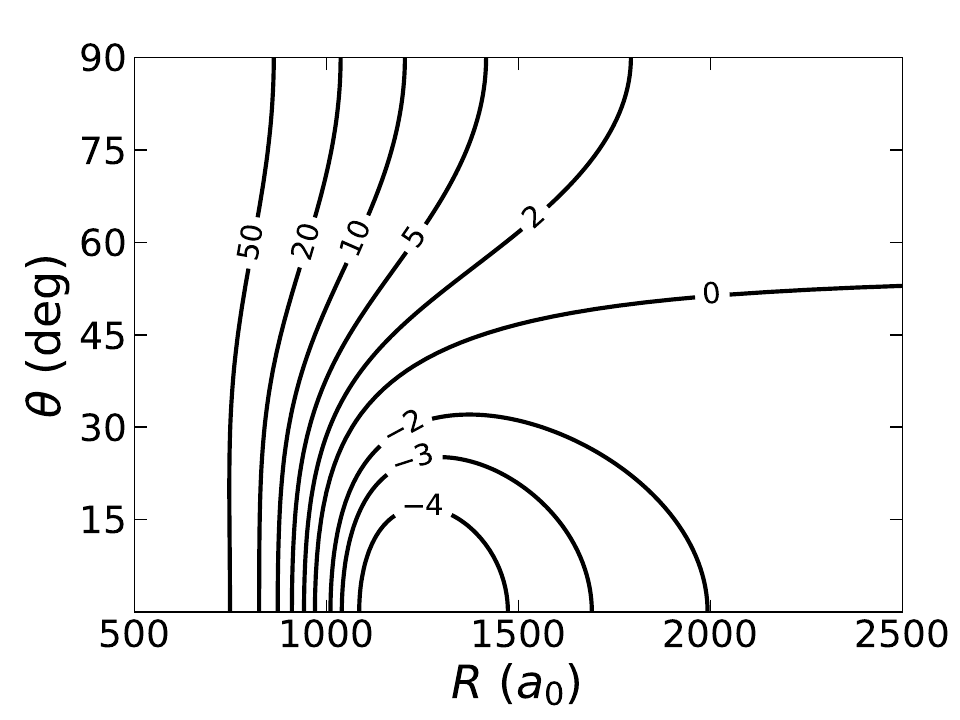}
    \caption{Contour plot of effective potential for CaF with double microwave shielding, with conditions as in Fig.\ \ref{fig:contour_MW}, supplemented with linearly polarised microwaves at Rabi frequency 30 MHz and detuning 4 MHz, using the functional form of ref.\ \cite{Deng:double-microwave:2025}. Contours are labeled in $\mu$K and shown only up to 50 $\mu$K.}
    \label{fig:contour_double}
\end{center}
\end{figure}

The effective potentials for static-field shielding may be contrasted with those for microwave shielding with a circularly polarised field \cite{Deng:microwave:2023}. The range and depth of the latter depend strongly on the microwave Rabi frequency and detuning. An example is shown in Fig.\ \ref{fig:contour_MW} for conditions that are reasonable for CaF. The potential is attractive at long range for angles $\theta_0<\theta<180^\circ-\theta_0$, but repulsive at all distances for $\theta<\theta_0$. This is opposite to the situation for static-field shielding. Both static-field and microwave shielding provide large repulsive cores that are very different from the short-range (contact-like) interactions that are common in atomic systems. However, the dipole-dipole forces resulting from static-field shielding favour geometries around $\theta=0$, while those from microwave shielding with circular polarisation favour geometries around $\theta=90^\circ$.

Interaction potentials with minima around $\theta=0$ can also be produced by double microwave shielding, combining circularly and linearly polarised microwave fields \cite{Karman:double:2025, Deng:double-microwave:2025}. In this case, however, shielding is maintained only when circular polarisation dominates. Strong losses occur with linear polarisation alone \cite{Karman:shielding:2018}, and this limits the strength of dipolar (as opposed to antidipolar) interactions that can be achieved with microwave shielding. Fig.\ \ref{fig:contour_double} shows an example of the dipolar interaction potential achievable with double microwave shielding for CaF. The potential is considerably weaker and has a different shape to that obtained with static-field shielding. Interaction potentials with minima around a single direction in space can also be produced by shielding with elliptical polarisation \cite{Chen:field-linked-resonances:2023}. However, the resulting interaction potentials are noncylindrical and depend on a large parameter space of microwave parameters, making detailed comparisons complicated. Moreover, elliptical polarisation causes enhanced losses \cite{Karman:shielding-imp:2019}.

A further difference between microwave and static-field shielding is that microwave shielding is near-universal, in the sense that the effective potentials are similar for all molecules when expressed in terms of the dipole length $R_3$ and energy $E_3$ \cite{Dutta:universal:2025}. With static fields, by contrast, the effective potentials depend strongly on $\tilde{b}$, as described in Sec.\ \ref{ssec:dcfield}: they are deepest and strongest (in reduced units) for molecules with large $\tilde{b}$ \cite{Mukherjee:alkali:2024}.

Effective potentials with attractive interactions around $\theta=0$ and $90^\circ$ will result in markedly different many-body physics. This will be true both for Bose-Einstein condensates of polar molecules and for polar molecules in optical lattices or tweezer arrays. For a sufficiently large number of molecules and strong-enough Rabi coupling, attraction around $90^\circ$ is expected to result in a dense self-bound disc-like droplet~\cite{Langen:dipolar-droplets:2025, Jin:Bose:2025}. For large couplings, this may eventually undergo a transition into a crystalline monolayer of molecules~\cite{Ciardi:self:2025}. In contrast, molecular condensates with attraction around $\theta=0$ should (under proper conditions) form elongated filament-like droplets aligned along the direction of the field. These droplets, though resembling those observed in magnetic atoms~\cite{Kadau:2016}, should not demand the stabilizing effect of quantum fluctuations. The properties of these self-bound droplets, including their stability, condensate fraction, and two-body correlations, are likely to be significantly different from those formed in the presence of microwave shielding. They may also present equilibrium multi-droplet solutions, which are not expected when the attraction is localised around $90^\circ$~\cite{Langen:dipolar-droplets:2025, Ciardi:self:2025}. This will open intriguing questions about supersolidity~\cite{Bottcher:2021} and pattern formation~\cite{Hertkorn:2021, Zhang:2021, Schmidt:2022}.

\section{Conclusions}

Ultracold polar molecules open up new frontiers for many-body physics. They provide experimentally realizable systems with strong dipole-dipole interactions, which challenge theories of dipolar quantum systems. However, ultracold molecules are stable only when protected from destructive short-range collisions by shielding methods. The interaction potentials offered by shielded ultracold molecules are different from the ones in common models. In particular, shielding produces interaction potentials with large repulsive cores, corresponding to repulsion at distances around 1000 bohr. This is very different from magnetic atoms, where the repulsion occurs at distances around 10 bohr and is fairly well represented by a contact (zero-range) interaction.

Simulations of condensed-phase or many-body behavior are computationally expensive. They require inexpensive ways to evaluate interaction potentials for each configuration. It is not computationally practical to diagonalize a substantial matrix at every point.

In this paper, we have examined the effective interaction potentials for pairs of ultracold molecules shielded from lossy collisions by static electric fields. We have derived forms of the effective potentials that can be evaluated efficiently and used in simulations of many-body properties. We have compared our interaction potentials with previous ones that gave no attraction or repulsion near $\theta=55^\circ$, and also with the ones appropriate for microwave shielding.

There are important contrasts between different types of shielding. Microwave shielding with circularly polarized fields gives effective potentials that are attractive for side-by-side geometries and favor molecules arranged in two-dimensional sheets. Static-field shielding, by contrast, gives effective potentials that are attractive for end-to-end geometries and favor molecules in linear arrangements. Microwave shielding with combined circular and linear polarisations can also give effective potentials that favour linear arrangements, but with important differences in strength and shape. Microwave and static-field shielding are complementary, and both are needed to explore the full range of many-body behavior that can exist.

\section*{Rights retention statement}

For the purpose of open access, the authors have applied a Creative Commons Attribution (CC BY) licence to any Author Accepted Manuscript version arising from this submission.

\section*{Data availability statement}

The data that support the findings of this article are openly available at https://doi.org/10.5281/zenodo.16966525.
A subroutine that evaluates the effective potentials efficiently is available from the authors on request.

\section*{Acknowledgement}
We are grateful to Michael Tarbutt and Joy Dutta for valuable discussions.
This work was supported by the U.K. Engineering and Physical Sciences Research Council (EPSRC) Grant Nos.\ EP/P01058X/1 and EP/V011677/1. L.S acknowledges the support of the Deutsche Forschungsgemeinschaft (DFG, German Research Foundation) -- Project-ID 274200144 -- SFB 1227 DQ-mat within the project A04, and under Germany's Excellence Strategy -- EXC-2123 Quantum-Frontiers -- 390837967).

\newpage
\bibliographystyle{../../long_bib}
\bibliography{../../all}

\end{document}